# On the physical nature of pseudogap phase and anomalous transfer of spectral weight in underdoped cuprates

Kirill Mitsen[1] and Olga Ivanenko

Lebedev Physical Institute, Leninsky pr., 53 Moscow, 119991, Russia

Abstract

It is shown that many anomalies observed in underdoped cuprates, including anomalous spectral weight transfer and a large pseudogap, appear to have a common nature due to both the cluster structure of the underdoped phase and the specific mechanism of superconducting pairing. The combined action of these factors leads to the fact that at a temperature T lying in a certain temperature range $T_c<T<T^*$, the crystal contains small isolated clusters that can exist both in superconducting and normal states, randomly switching between them. In this case, below Tc with a very high probability the cluster is in a superconducting state, and above T* it is in a normal state, and the interval $T_c<T<T^*$ is the region of existence of the so-called pseudogap phase. The temperatures $T_c$ and T* for $YBa_2Cu_3O_{6+\delta}$ were calculated depending on the doping level $\delta$. The calculation results are in good agreement with experiment without the use of fitting parameters. At a given T in the same temperature range, the time sequence of randomly arising superfluid density pulses from each cluster can be represented as a random process. The effective width $\Delta\omega_{eff}$ of the spectrum of such a random process will be determined by a correlation time, i.e. the characteristic time between successive on/off superconductivity in two different clusters. This time, according to the estimate, is $\sim 10^{-15}$ sec, which corresponds to $\Delta\omega_{eff} \sim 1$ eV and explains the effect of spectral weight transfer to the high-frequency region. This approach also makes it possible to explain other anomalies observed in the vicinity of Tc: the reversibility of magnetization curves in a certain temperature range below Tc, the anomalous Nernst effect and anomalous diamagnetism above Tc.

To date, a huge amount of experimental material has been accumulated on the observation of various anomalies in underdoped cuprates, which, as a rule, are associated with the existence of a pseudogap. However, despite many experiments performed, the underlying physics of the pseudogap is still shrouded in mystery.

Analyzing the results of these experiments, one can see that the concept of a pseudogap is used, in fact, to describe two different phenomena. The first is the existence of an energy gap on part of the Fermi surface in a certain temperature range above Tc. This pseudogap is detected directly by angle-resolved photoemission spectroscopy (ARPES) [1,2,3,4] and STM [5] which can measure the density of states of the electrons in a material. The pseudogap observed in these experiments is usually called a "small" pseudogap and its magnitude decreases monotonically with increasing hole concentration (doping). An explanation of the nature of this pseudogap based on the model we propose is given in [6].

In addition to the "small" pseudogap, a "large" pseudogap is also considered, which appears, although more indirectly, in NMR measurements of $YBa_2Cu_3O_{6+x}$ [7], in heat capacity measurements [8], in resistive [9,10] and optical [11] measurements etc. It manifests itself in the form of various "superconducting" anomalies below the temperature T*(> Tc) increased with doping decrease. Another mysterious phenomenon, especially manifested in underdoped cuprates, is spectral weight (SW) transfer induced by the superconducting transition [12]. In overdoped cuprates, the superfluid response develops due to states lying below 100 meV, i.e. energies of the order of several superconducting gaps, as in conventional superconductors. In underdoped cuprates, a decrease in spectral weight of up to 1 eV is observed, which is significantly larger than any conventional scale. The existence of a pseudogap phase and anomalous SW transfer indicate that either the pairing mechanism in underdoped cuprates is very unconventional, or these anomalies are associated with the special nature of the underdoped state.

---

[1] Correspondence to [mitsen@lebedev.ru]

We have previously shown that, due to local deformation of the electronic structure of cuprates and pnictides by impurity ions, the doping process at low concentrations (in underdoped region) is accompanied by the formation of individual clusters of a new phase with a special electronic structure [13], in which a specific mechanism of superconductivity may be realized [14]. We have shown that the positions of the superconducting domes in the phase diagrams of cuprates and pnictides coincide with the existence intervals of percolation cluster of this phase. This fact suggests that high-temperature superconductivity takes place just in this new phase [13,15].

Here we will consider the peculiar effect that can arise in the underdoped phase when taking into account its cluster structure and the proposed mechanism of superconductivity. This effect is the possibility of finite clusters to exist in a certain temperature range in both superconducting and normal states, switching between these states randomly. This effect allows us to give a unified explanation for many anomalies of the underdoped phase, including the existence of a pseudogap phase, anomalous SW transfer, reversibility of magnetization curves in a certain temperature region below Tg, anomalous diamagnetism above Tg and the anomalous Nernst effect.

## Anomalous transfer of spectral weight

As is known, for a system of interacting electrons the following rule is valid (the f-sum rule, or Thomas-Reich-Kuhn rule), obtained by R. Kubo [16], for the real part of the complex optical conductivity $\hat{\sigma}(\omega) = \sigma_1(\omega)+i\sigma_2(\omega)$ [17].

$$\int_0^\infty \sigma_1(\omega)d\omega = \pi n e^2/(2m) \qquad (1)$$

where *n,e* and *m* are the electron density, charge, and mass, respectively. It follows that at all temperatures the total spectral weight (SW) is constant and depends only on the electron density in the system. The functional dependence **σ₁(ω)** can change, for example, with temperature, and the full spectral weight is preserved and is only redistributed between different frequencies as the temperature changes. For single-band materials, taking into account only intraband transitions, we can limit the integration range to the frequency $\Omega_m \sim W/\hbar$, where *W* is the band width.

In superconductors, the transition to the superconducting state is accompanied by the opening of the energy gap $2\Delta_0$ and the appearance of $\delta$ function at zero frequency in the conductivity, which describes the superfluid response. The spectral weight of this $\delta$ function is $D=\pi e^2(n_s/2m^*)$ [18], where $n_s$ is the concentration of superconducting electrons, $m^*$ is the effective mass of electron. Thus, the optical conductivity of a superconductor $\sigma_{1s}(\omega)$ has both a regular part $\sigma_{1s}^{reg}(\omega)$ (at ω>0) and a singular part $D\delta(\omega)$ and can be written in "two-fluid" form.

$$\sigma_{1s}(\omega)=D\delta(\omega) + \sigma_{1s}^{reg}(\omega) \qquad (2)$$

Hence the spectral weight of this $\delta$ function

$$D = \int_{0+}^\infty [\sigma_{1,n}(\omega) - \sigma_{1,s}(\omega)]d\omega \equiv \delta A, \qquad (3)$$

where $\sigma_{1,n}(\omega)$ is the conductivity in the normal state, and $\delta A$ is the "missing area" under the curve $\sigma_{1n}(\omega)$ at the transition to the superconducting state (Fig. 1).

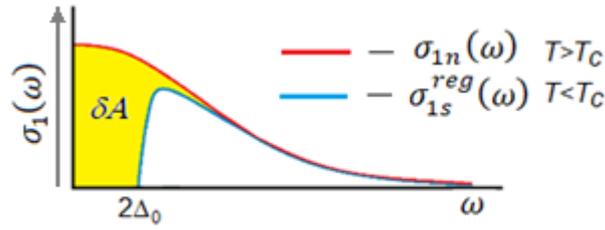

Fig. 1. The optical conductivities of superconductor in the normal state - $\sigma_{1n}(\omega)$ and SC state - $\sigma_{1s}^{reg}(\omega)$ at ω> 0. Gray arrow - $\delta$ peak at zero frequency.

Superfluid weight *D* also manifests itself at the finite frequencies through a Kramers-Kronig relation and gives rise to a 1/ω contribution to the imaginary part of the conductivity σ₂(ω) that can be measured independently. Fig. 1 shows typical frequency dependencies $\sigma_{1n}(\omega)$ (red curve) and $\sigma_{1s}^{reg}(\omega)$ (blue curve) for conventional superconductor. The transition to a superconductive state, accompanied by the opening of the energy gap 2$\Delta_0$, manifests itself (Fig. 1a) as suppression $\sigma_{1n}(\omega)$ for frequencies in the area $\hbar\Omega_c \sim 4\Delta_0$ (Gap Region). The decrease in spectral weight in this area $\delta A$, (painted yellow in Fig. 1a) is quite accurately compensated by the appearance of the deposit *D* from the conductivity at zero frequency .
For superfluid density $\rho_S = c^2/\lambda^2 = 4\pi e^2 n_s/m^*$, using Cubo sum rules, we can get the ratio

$$\rho_s = 8 \int_{0+}^{\infty} [\sigma_{1n}(\omega) - \sigma_{1s}^{reg}(\omega)] d\omega, \tag{4}$$

In conventional superconductors, the SW from the superfluid accumulates from *ω < Ωc* frequencies, so integration in (3) can be carried out over a relatively narrow frequency range.

Unlike conventional superconductors, in underdoped cuprates SW from the near-gap region within the energy scale $\hbar\Omega_c \sim$ 0.15 eV supply only a part of the superconducting condensate and one should suggest the SW transfer at superconducting transition from much higher energies, up to the near-infrared region of 0.5–1 eV. This SW deficit was observed when measuring both along the ab-plane [19,20,21,22 ] and along the c-axis [12,24,25,26 ]. The results obtained are usually explained by a decrease in kinetic energy at the superconducting transition, which distinguishes underdoped cuprates from conventional superconductors, where the transition to the superconducting state is accompanied by an increase in kinetic energy. This explanation is based on the results of Maldague [27], where the "partial" sum rule was derived within the framework of the single-zone Hubbard model.

$$\int_0^{\Omega_0} \sigma_1^{(r)}(\omega) d\omega = \frac{\pi e^2 a_r^2}{2\hbar^2} K_r \tag{5}$$

Here $a_r$ and $K_r$ are, correspondingly, the lattice spacing and the electronic kinetic energy in the *r* direction, and $\Omega_0$ is a cutoff frequency which restricts the allowed transitions to states inside the single band. This makes it possible to determine the contribution to the density of the superconducting condensate from the frequency range exceeding Ω₀ with a change (decrease) in the kinetic energy in the r direction during the superconducting transition [28].

$$D_r = \int_{0+}^{\Omega_0} \sigma_{1,n}^{(r)}(\omega) - \sigma_{1s}^{(r)}(\omega)] d\omega + \frac{\pi e^2 a_r^2}{2\hbar^2} \Delta K_r \tag{6}$$

In anisotropic cuprates, two possible types of kinetic energy should be distinguished: perpendicular to the planes (along the c-direction) and along the planar directions [29]. Accordingly, the measurements of the optical conductivity along the c-axis [30] as well as along the copper-oxide planes [31, 32, 33] have been used to estimate the changes in kinetic energy in both directions. The experiments have indicated that, an influence of superconductivity on the optical spectrum exists at frequencies up to the near-infrared region of 0.5–1 eV for light polarized both in *ab* plane and along the c-axis. We emphasize that the effect of SW deficiency or a decrease in kinetic energy at the superconducting transition is most clearly observed in underdoped cuprates, while overdoped cuprates behave like conventional superconductors and do not exhibit a spectral weight deficit at superconducting transition [25]. Moreover, even in underdoped cuprates, this effect disappears in magnetic fields B>8T [34].

The noted anomalies in the optical properties of cuprates represent a significant departure from the BCS model and suggest that either the HTSC mechanisms are different for different regions on the phase diagram, or, if this mechanism is general, it should explain a transition from a decrease in kinetic energy to its increase as doping increases. However, this contradiction is removed if we assume that the dependence $\sigma_{1n}(\omega)$, relative to which changes due to the superconducting transition are measured, is in fact not the true optical conductivity of the sample in the normal state.

Specifically, a sample of underdoped cuprate HTSC with a critical temperature $T_c<T_{c\infty}$ ($T_{c\infty}$ is the temperature of the superconducting transition of an optimally doped superconductor), being at a measurement temperature T slightly above $T_c$, is not, strictly speaking, homogeneous and completely normal, but contains a large number of small clusters, each of which, in a certain temperature range $T_c<T_{c0}<T^*$, can exist in both superconducting and normal states, randomly switching between these states with a characteristic switching time $\tau_{SN}$~$\hbar/2\Delta$ ~$10^{-14}$ sec. These clusters can combine into short-lived Josephson-coupled superconducting clusters of larger size. We will discuss the nature of such "flickering" clusters later.

In the case of a large number of small clusters, we can represent the contribution to optical conductivity from a sequence of fluctuationally arising superfluid density pulses from many individual clusters as a random process. The effective width $\Delta\omega_{eff}$ of the spectrum of a random process $S(\omega)$ is determined as $\Delta\omega_{eff} = \frac{1}{S_{max}}\int_0^\infty S(\omega)d\omega$. A graphical interpretation of the effective spectrum width of a random process is shown in Fig. 2.

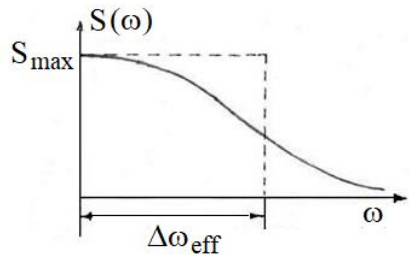

Fig. 2 The graphical interpretation of the effective spectrum width of a random process

The effective spectrum width of a random process $\Delta\omega_{eff}$ is related to the correlation interval $\tau_{corr}$ by the relation [35]:

$$\Delta\omega_{eff}\ \tau_{corr} > \pi/2 \qquad (7)$$

The correlation interval shows the time of probable forecast of a random process. Forecasting beyond the correlation interval is not possible. For the random process under consideration, the correlation interval $\tau_{corr}$ is determined by the characteristic time between successive on/off of superconductivity in two random clusters. This time $\tau_{corr}$ may be shorter than $\tau_{SN}$. Based on the fact that in the experiment SW transfer is observed in the range up to 1 eV, it should be assumed that $\tau_{corr}$~$10^{-15}$sec. As will be shown below, it is just this value $\tau_{corr}$ that characterizes random process under consideration. This time corresponds to the transition of an electron between oxygen and copper ions, which, under certain conditions, can be a trigger for switching of cluster between the superconducting and normal states.

Let us consider how the optical conductivity spectrum of a conventional superconductor with the superconducting transition temperature $T_c$ (Fig. 3a) should change at transition to an inhomogeneous superconductor in which there are small clusters with the superconducting transition temperatures $T'_c$ ($T_c < T'_c < T^*$) (Fig. 3b). On this figure: the red curve is $\sigma_{1n}(\omega)$ for T>T*, the blue curve – $\sigma_{1s}^{reg}(\omega)$ for T<$T_c$; brown curve – optical conductivity at $T_c$<T<T*; the spectral weight from small clusters transferred to the δ-peak at ω=0 is colored yellow. If we assume that small clusters fluctuate with a high frequency between the superconducting and normal states (with the effective spectral width of the random process Δω$_{eff}$), then the spectral weight shaded in Fig. 3b in yellow, will no longer be transferred to the δ-peak, but should, in accordance with Fig. 2, redistribute over the interval $\Delta\omega_{eff} \gg 2\Delta$. As a result, in inhomogeneous superconductor the dependence $\sigma_{1n}^{inh}(\omega)$ will take the form shown in Fig. 3c with black line. Comparing this dependence $\sigma_{1n}^{inh}(\omega)$ with $\sigma_{1s}(\omega)$, we can erroneously conclude that states from the entire $\Delta\omega_{eff}$ region (shaded in green) contribute to the spectral weight of the δ-peak. We emphasize that this effect is most pronounced in underdoped cuprates, which have a cluster structure of the superconducting phase. In optimally doped cuprates in which superconductivity occurs in a percolation cluster, this effect is much weaker and is due to only a small number of remaining small clusters. Overdoped samples, in which the superconducting cluster occupies the entire $CuO_2$ plane [13-15], will behave like ordinary superconductors.

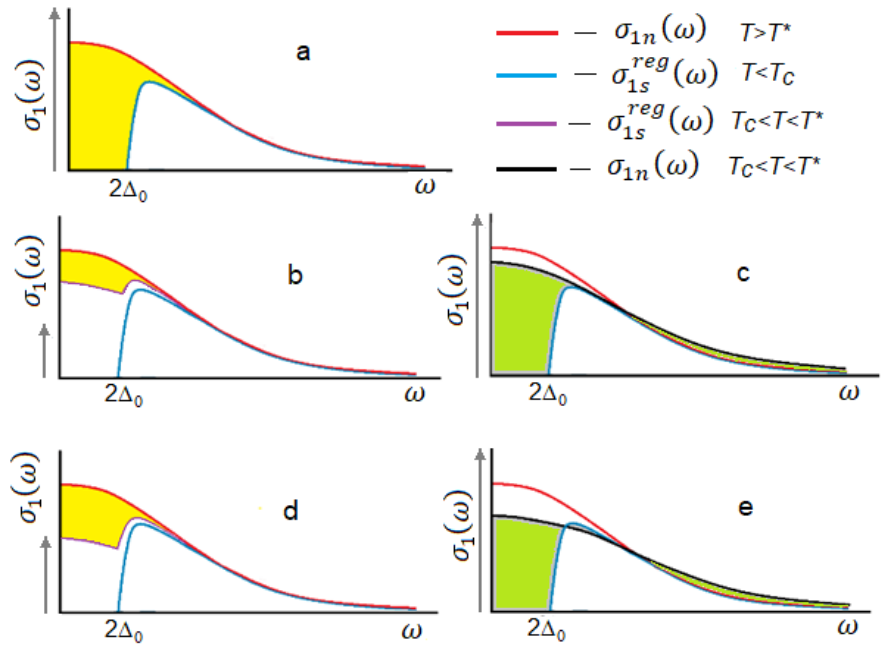

Fig. 3 (a) Optical conductivity spectrum of a spatially homogeneous superconductor: red curve - $\sigma_{1n}(\omega)$, blue curve – $\sigma_{1s}^{reg}(\omega)$; the spectral weight transferred to the δ-peak (grey arrow) at ω=0 is shaded yellow. (b) Hypothetical spectrum of optical conductivity of a spatially inhomogeneous superconductor (with a temperature of bulk superconductivity $T_c$), including small clusters with a transition temperature $T'_c$ ($T_c < T'_c < T^*$); red curve – $\sigma_{1n}(\omega)$ at T>T*, blue curve – $\sigma_{1s}^{reg}(\omega)$ at T<$T_c$; brown curve – optical conductivity at $T_c$<T<T*; the spectral weight from small clusters transferred to the δ-peak at ω=0 (gray arrow) is colored yellow; (c) red curve – $\sigma_{1n}(\omega)$ at T>T*, when the entire sample is in a normal state; blue curve – $\sigma_{1s}^{reg}(\omega)$ at T<$T_c$, when the entire sample is in a superconducting state; black curve is $\sigma_{1n}^{inh}(\omega)$- optical conductivity spectrum at $T_c$<T<T* of a inhomogeneous superconductor, including small clusters fluctuating between the superconducting and normal states, gray arrow - δ-peak at T<$T_c$; (d) and (e) are the same as (b) and (c), respectively, but with a larger number of small clusters. The superconductivity-induced change of the conductivity ($\sigma_{1n}^{inh}(\omega) - \sigma_{1s}^{reg}(\omega)$) reverses sign in the region of maximum $\sigma_{1s}^{reg}(\omega)$.

This explanation also allows us to understand the more subtle features of the conductivity changes caused by superconductivity. In particular, the sign-alternating character of the change observed in the optical conductivity along the c axis of $La_{2-x}Sr_xCuO_4$ [36,37]. Actually, let consider the case where the number of small clusters is large (Fig. 3d compared to Fig. 3b). In this case, more spectral weight will be smeared over the interval $\Delta\omega_{eff}$ and the value of $\sigma_{1n}^{inh}(\omega)$ in some region $\omega > 2\Delta_0$ will be lower than $\sigma_{1s}^{reg}$. This will correspond to the positive change of the low-frequency spectral weight in the region of the maximum of $\sigma_{1s}^{reg}$, which was observed in the experiment [36,37]. Also, with this approach, the suppression of anomalous SW transfer by a magnetic field [34] can be explained, since the magnetic field destroys superconductivity in finite clusters at $T>T_c$. Thus, assumptions about the structure of the underdoped phase of cuprates as a medium consisting of isolated finite clusters oscillating in a certain temperature range between the superconducting and normal states make it possible to explain the main features of the optical response observed in numerous experiments. Next, within the framework of the previously proposed model [13-15], we consider the nature of the cluster structure of underdoped cuprates, and what mechanism of superconducting pairing allows small clusters to switch at a constant temperature between normal and superconducting states.

## The nature of cluster structure of underdoped cuprates

The proposed model of heterovalently doped cuprates [13-15] is based on the assumption of the local character of the electronic structure deformation of the crystal by the dopant ion. We consider initially undoped cuprates as charge-transfer (CT) insulators with a gap $\Delta_{ct}$ [38]. The local nature of the deformation of their electronic structure is determined by the localization of the doped carrier in the immediate vicinity of the dopant due to the formation around the doped charge of a trion complex - the bound state of the doped carrier and CT excitons that arise under the influence of this charge in the surrounding $CuO_4$ plaquettes (we called such plaquettes as CT plaquettes) [13,14]. This becomes possible if the doped charge suppresses the initial charge-transfer gap $\Delta_{ct}$ in neighboring $CuO_4$ plaquettes, to a value $\Delta'_{ct} < \Delta_{ct}$, sufficient for the production of CT excitons in these plaquettes [13]. Note that the interaction of such trion complexes corresponds to repulsion, which contribute to their ordering.

When dopant concentration increases CT plaquettes formed in the $CuO_2$ plane in the vicinity of each dopant ion are combined into plaquette clusters (CT clusters). We will call a CT cluster a finite CT cluster if it is a collection of CT plaquettes completely surrounded by $CuO_4$ plaquettes that are not CT plaquettes According to the model [14], superconductivity in cuprates arises just in CT clusters:

The geometry of the arrangement of trion complexes and associated CT plaquettes is determined by the position of dopants in the crystal lattice and is generally different for each compound. It was shown [13,15] that most of the anomalies in the superconducting characteristics of cuprates observed at dopant concentrations inside the superconducting dome, as well as the very position of the domes on the phase diagrams of compounds, do not require precise knowledge of their electronic structure for explanation, but can be understood and calculated with high accuracy within the framework of a simple geometric model describing the cluster structure of the superconducting phase.

According to the model, the underdoped phase corresponds to a range of dopant concentrations below the percolation threshold for CT plaquettes in the $CuO_2$ plane, when the finite CT clusters are immersed in an insulating matrix, and superconductivity in the entire sample arises due to the Josephson coupling between the clusters. In this case, the region of optimal doping corresponds to the existence of a percolation cluster of CT plaquettes. In the overdoped phase, the distance between neighboring doped charges decreases in such a way that in the space between them the gap $\Delta_{ct}$ = 0 and the remaining finite clusters of CT plaquettes are immersed in the normal matrix. In the underdoped (overdoped) phase, the average cluster size decreases with decreasing (increasing) doping.

## Heitler-London centers and a possible mechanism of superconductivity

The proposed mechanism of local modification of the electronic structure of cuprates upon doping also suggests the possibility of the formation of bound electronic states on pairs of Cu ions in adjacent CT plaques [14]. We called such a hydrogen-like pair of CT plaquettes the Heitler-London center (HL center). Thus, in a CT cluster, each pair of neighboring CT plaquettes is an HL center, and the CT cluster itself is a continuous network of HL centers. The generation of hole carriers in the system occurs due to the transfer of pairs of electrons from the oxygen band to pairs of neighboring Cu ions in HL centers [14]. It is these holes that are free carriers, in contrast to doped carriers, which are localized in the vicinity of the dopant. The concentrations of free holes and localized electron pairs at HL centers depend on temperature. The existence of HL centers, in addition to the generation of carriers, also provides the possibility of implementing a specific mechanism of superconducting pairing [14], which originates from virtual transitions of band electrons into unfilled pair states and resembles of the Little-Ginsburg mechanism [39,40].

Thus, as follows from the proposed model, HL centers play a dual role: they act as a kind of "acceptors", the filling of which with real electrons leads to the generation of free carriers in the system, and they are also centers of superconducting pairing. Note that the more pair states in HL centers are filled with real electrons, the less HL centers remain available for virtual transitions of electron pairs, which means the pairing potential decreases. Thus, the parameter that directly controls the transition to the superconducting state is the coefficient of filling of HL centers.

In this case, a random increase in the occupation of pair states above the equilibrium value will reduce the pairing interaction and can lead to the destruction of superconductivity in a separate cluster, starting from the temperature $T_c<T_{c\infty}$ (here $T_{c\infty}$ is the temperature of the superconducting transition of an infinite cluster, for which the relative occupation fluctuations are negligible ), and, conversely, a decrease in the occupation of pair states below the equilibrium value can lead to the emergence of superconductivity in this cluster, starting from $T^*>T_{c\infty}$. Thus, in a certain temperature range $T_c<T<T^*$, a cluster can exist in both superconducting and normal states, randomly switching between these states as a result of fluctuations in the occupation of HL centers. This effect of s-n switching will be most pronounced in small clusters containing a small number of pair states (or including a small number of Cu ions), when the relative fluctuations in the occupation of pair states will be large.

Returning to the problem of SW transfer, let us consider how flashing of superfluid density at frequent switching between superconducting and normal states affect the optical conductivity spectrum $S(\omega)$. Since the state of the cluster is determined by the occupation of pair states, to change the state of small cluster it is necessary to transfer one or more pairs of electrons from the oxygen band to the pair states of HL centers or back. In this case, the correlation interval (or the characteristic time between successive switchings of two clusters) $\tau_{corr}$ can be estimated as the time of transition of a pair of electrons to the pair states of HL centers or back - $\tau_{corr} \sim a/2v_F \sim 10^{-15}$ sec. Here $a$ is the lattice constant in the $CuO_2$ plane, and $v_F$ is the Fermi velocity. In accordance with (7), the effective spectrum width of such a process is $\Delta\omega \sim 10^{15}$ sec$^{-1}$. Thus, if at the transition of the entire sample to the superconducting state, the spectral weight of states from the region $\hbar\Omega_c \sim 4\Delta_0$ goes to the δ-peak at $\omega=0$, then in the case of the existence of small clusters fluctuating between normal and superconducting states, the spectral weight is redistributed over a wide spectral range up to energies of ~1 eV.

Accordingly, for a sample, containing small clusters the optical conductivity spectrum $\sigma_{1,n}(\omega)$ in the interval $T_c<T<T^*$ changes relative to the spectrum of an infinite cluster. The resulting changes will correspond to a decrease in $\sigma_{1,n}(\omega)$ in the low-frequency region $\hbar\Omega_c \sim 4\Delta_0$ and an increase in the high-frequency region up to ~ 1 eV, that creates the illusion of an anomalous SW transfer from the high-frequency region to the δ-peak.

# Calculation of $T_c$ and $T^*$ dependences on doping for $YBa_2Cu_3O_{6+\delta}$

Now, using $YBa_2Cu_3O_{6+\delta}$ as an example, we evaluate the possibility of observing such fluctuations and their scale. For this purpose, we determine the average size of a CT cluster depending on the doping level $\delta$ and calculate the characteristic temperatures $T_c$ and $T^*$, in the interval between which a cluster of a given size can switch between superconducting and normal states.

According to the model [13, 15], CT plaquettes in $CuO_2$ planes $YBa_2Cu_3O_{6+\delta}$ are formed when two adjacent oxygen sites in the chain are filled (Fig. 4 a). Such CT plaquettes are formed in each of the two $CuO_2$ planes, adjacent to the CuO chain. In this case, doped oxygen ions induce positive charges q~e/4 on the apical ions, which initiate the creation of CT excitons in the nearest $CuO_4$ plaquettes, leading to the formation of a trion complex, i.e. bound state of the doped charge and two surrounding $CuO_4$ plaquettes. The size of a CT cluster in $CuO_2$ plane, consisting of a certain number of CT plaquettes, will be measured by the number of Cu sites that are the centers of these plaquettes. As mentioned above, in a CT cluster, each pair of CT plaquettes which centered on neighboring Cu ions in the $CuO_2$ plane forms a hydrogen-like HL center (Fig. 4b).

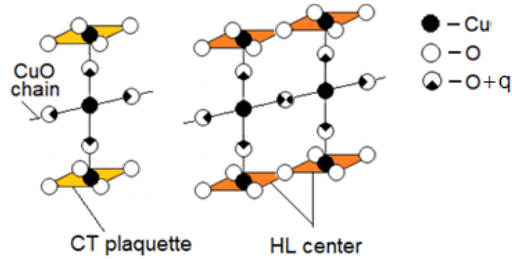

Fig. 4 (a) CT plaquettes in $CuO_2$ planes of $YBa_2Cu_3O_{6+\delta}$ are formed when two adjacent oxygen positions in the chain are filled. (b) a pair of CT plaquettes which centered on neighboring Cu ions in the $CuO_2$ plane forms a hydrogen-like HL center.

Since the formation of a CT plaquette requires the filling of two adjacent oxygen positions in the chain, the number of such CT plaquettes in the $CuO_2$ plane is $\delta^2$ (i.e., the probability $p$ that a given Cu site is the center of a CT plaquette is $\delta^2$). The percolation threshold for sites problem in a square lattice is $p_{th}$= 0.593 [41]. Therefore, for the formation of a percolation cluster of CT plaquettes, the condition $\delta^2_{th}$=0.593 must be satisfied. Hence $\delta_{th}$=0.77, which coincides with the experimental boundary of the optimal doping region [42,43].

When the oxygen concentration is below the percolation threshold $\delta_{th}$=0.77, the percolation cluster of CT plaquettes disintegrates into finite CT clusters. Accordingly, at $\delta$>0.77 in the $CuO_2$ plane there is a percolation HL cluster. In $YBa_2Cu_3O_7$, all Cu ions in the $CuO_2$ plane belong to HL centers, and all $CuO_2$ plane is a percolation HL cluster. In an HL cluster (finite or percolation), due to electronic transitions between the states of the oxygen band and localized paired states of HL centers, a dynamic equilibrium is established with a common chemical potential μ.

For small hole concentrations the change of the chemical potential for electron pairs $\Delta\mu_p$ due to transition of part of electrons from a band to HL centres can be presented as [6,44]:

$$\Delta\mu_p = -\frac{T}{2}ln[f(n)] \qquad (8)$$

where $n$ is the two-dimensional concentration of formed holes per Cu ion, and $f(n)$ is some function of occupation of HL centres. Herewith, the change of the chemical potential for band electrons $\Delta\mu_e \approx -n/N(0)$, where $N(0)$ is the density of states on the Fermi level in a band. Equating the chemical potential changes of pairs and band electrons, we find in an explicit form the dependence of hole concentrations $n$ on temperature:

$$T \approx 2n/N(0)ln[f(n)] \tag{9}$$

Thus at low *T*, $n \propto T$. At high *T*, the concentration tends to a saturation, when the coefficient of filling of copper sites tends to unity (i.e., all HL centres are occupied). The entire dependence can be approximated by the formula

$$n = T/(T + T_0), \tag{10}$$

where $T_0$ is a certain constant that does not depend on temperature. As shown in [45], such a dependence (with $T_0$=390K) perfectly describes the doping and temperature dependences of the measured Hall effect in $YBa_2Cu_3O_{6+x}$. Relation (10) simultaneously determines both the hole concentration and the occupation of HL centers, i.e. concentration of filled Cu sites. Note that for optimally doped $YBa_2Cu_3O_{6+x}$, where a percolation CT cluster exists, the superconducting transition temperature $T_{c0}$=93K corresponds to the concentration $n \approx 0.2$.

Below the percolation threshold $\delta$<0.77, the percolation cluster disintegrates into finite CT clusters, which are also HL center clusters. Their average size decreases rapidly with decreasing doping. Under these conditions, the influence of fluctuations of HL center occupation on the superconductivity of finite clusters increases sharply. Since the pairing potential depends on the concentration of unoccupied HL centers, an increase in the occupancy of HL centers above the equilibrium value due to fluctuations will reduce the pairing interaction and lead to the destruction of superconductivity in a separate cluster at $T<T_{c\infty}$, and, conversely, a decrease in the occupancy of pair states below the equilibrium value can lead to to the emergence of superconductivity in this cluster at $T>T_{c\infty}$. This effect will be most pronounced in small clusters containing a small number of pair states (or including a small number of Cu ions), when the relative fluctuations in the occupation of pair states will be large. With decreasing doping, the average size of the finite clusters decreases and the relative fluctuations in the occupation of HL centers in these clusters increase (i.e., T* will increase and $T_c$ will decrease). Based on the proposed model, it is possible to determine the dependences of T* and $T_c$ on the cluster size (and, accordingly, on the doping level δ). We will assume that at $\delta<\delta_{th}$, the sample is a Josephson medium, where superconductivity throughout the entire volume is achieved due to Josephson coupling between superconducting clusters. Let us recall that we take the cluster size - *S* as the number of Cu sites included in it.

Let us consider a CT cluster of S size in the $CuO_2$ plane, including a certain number of HL centers. According to (3), the equilibrium number N of electrons on Cu ions in a given cluster at temperature T is equal to $N(T)=TS/(T+T_0)$, where $T_0$=390K. We will assume that, as a result of fluctuations, the number of electrons on Cu ions in a given cluster has a Poisson distribution and can change by $\pm\sqrt{N}$. The condition for the fluctuation "switching off" ("switching on") of superconductivity in a given cluster at temperature $T_c$ (*T**) can be written as $N(T) \pm \sqrt{N(T)} = N_c$, where $N_c = T_{c\infty}S/(T_{c\infty}+T_0) \approx 0.2S$ is the equilibrium number of electrons on Cu ions in this cluster at $T=T_{c\infty}$. The minus sign corresponds to $T=T^*$, and the plus sign to $T=T_c$. Thus

$$\frac{TS}{T+T_0} \pm \sqrt{\frac{TS}{T+T_0}} = \frac{T_{c\infty}S}{T_{c\infty}+T_0} = 0.2S \tag{11}$$

Since fluctuations of the number of occupied Cu ions in a cluster must be integer, for small clusters at low temperatures these fluctuations should be considered as average fluctuations in the occupation of Cu ions included in CT clusters in the $CuO_2$ plane. Similarly, $T_c$, determined from (11), should be considered as the temperature of the superconducting transition of a Josephson coupled cluster uniting small superconducting clusters. Solving the quadratic equation (11), we find two roots T* and $T_c$ as functions of S (Fig. 5).

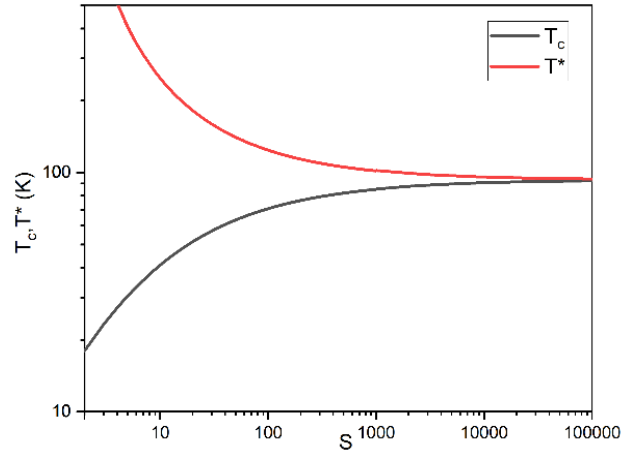

Fig. 5. The dependences of temperatures T* and $T_c$ for $YBa_2Cu_3O_{6+x}$ on cluster size S. Both temperatures obtained as two solutions of a quadratic equation without using fitting parameters.

Since in a sample with a concentration $\delta$<0.77 there are finite clusters of various sizes, in order to understand how the temperatures $T^*$ and $T_c$ behave for the entire sample depending on doping, we use the concept of a mean cluster size $\bar{S}$ (mean cluster size) and we will assume that the dependences $T^*(\delta)$ and $T_c(\delta)$ are determined by the dependence $\bar{S}(\delta)$.

Within the framework of the site percolation problem on the simple quadratic lattice, the average size of a CT cluster $\bar{S}(p)$ (i.e., the number of Cu sites that it includes) depending on doping is determined as:

$$\bar{S}(p) = \sum_{s=1} s^2 \langle n_s \rangle / \sum_{s=1} s \langle n_s \rangle \tag{12}$$

where $n_s$ is the number of clusters of size S. Note that we are talking about a weighted average here. For a simple square lattice, the value $\bar{S}$, depending on the probability of a given site filling – $p$, can be represented as a convergent series [46]

$$\bar{S}(p) = \sum a_n p^n = 1 + 4p + 12p^2 + 24p^3 + 52p^4 + 108p^5 + 224p^6 + 412p^7 + 844p^8 + 1528p^9 + 3152p^{10}\ldots \tag{13}$$

Substituting $p=\delta^2$ into this formula, we obtain $\bar{S}(\delta)$ - the doping dependence of the average size of a CT cluster (measured in the number of Cu ions included in it) in $YBa_2Cu_3O_{6+\delta}$ - (Fig.6, curve 1).

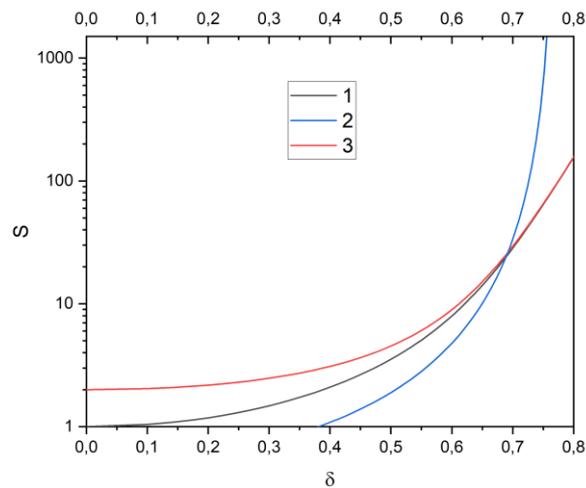

Fig. 6 Dependence of the average size of the CT cluster on doping $\bar{S}(\delta)$ for $YBa_2Cu_3O_{6+\delta}$. The percolation threshold for CT plaquettes corresponds to the concentration $\delta_{th}$ =0.77. Curve 1 – average size of CT cluster $\bar{S}(\delta)$ according to (13); 2 - average size of a CT cluster near the percolation threshold - $\bar{S}_{th}(\delta)$; 3 - average size of HL cluster – $\bar{S}2(\delta)$,

Expression (13) can only be used for *p* not too close to the percolation threshold $p_{th}$ =0.593 or $\delta_{th}$=0.77. In the region $\delta_{th}-\delta\ll\delta_{th}$ the relation (14) should be used [47]. This dependence is shown in Fig. 6 (curve 2).

$$\bar{S}_{th}(\delta) = 0{,}147\,(0{,}593-\delta_c^2)^{-43/18} \qquad (14)$$

As can be seen from Fig. 6 (curve 1) at a concentration $\delta<0.4$, the average weighted size of the CT cluster is $\bar{S}<2$, i.e. the most of clusters consist of a single CT plaquette and do not contain HL centers. Such CT clusters do not belong to a Josephson coupled superconducting cluster. The average size of an isolated cluster containing HL centers can be determined as

$$\bar{S}2(\delta) = \sum_{s=2} s^2 \langle n_s \rangle / \sum_{s=2} s \langle n_s \rangle \qquad (15)$$

where the summation is carried out starting from *s*=2. Only such clusters with s≥2 can form a large superconducting cluster, in which superconductivity arises due to Josephson coupling between small clusters. The dependence $\bar{S}2(\delta)$ is shown in Fig. 6 (curve 3).

Now, when we know the statistics of finite clusters depending on doping (Fig. 6), we can determine the dependences $T^*(\delta)$ and $T_c(\delta)$. For the entire range of $\delta$ we use the values $\bar{S}(\delta)=\bar{S}2(\delta)$ on the range $0<\delta<0.7$ and $\bar{S}(\delta)=\bar{S}_{th}(\delta)$ for $\delta>0.7$. The values of $T^*$ and $T_c$ are obtained as 2 solutions of one quadratic equation (11). A comparison of the calculated dependences $T^*(\delta)$ and $T_c(\delta)$ with the experimentally measured ones is shown in Fig. 7. Note that the calculated dependences do not contain any fitting parameters.

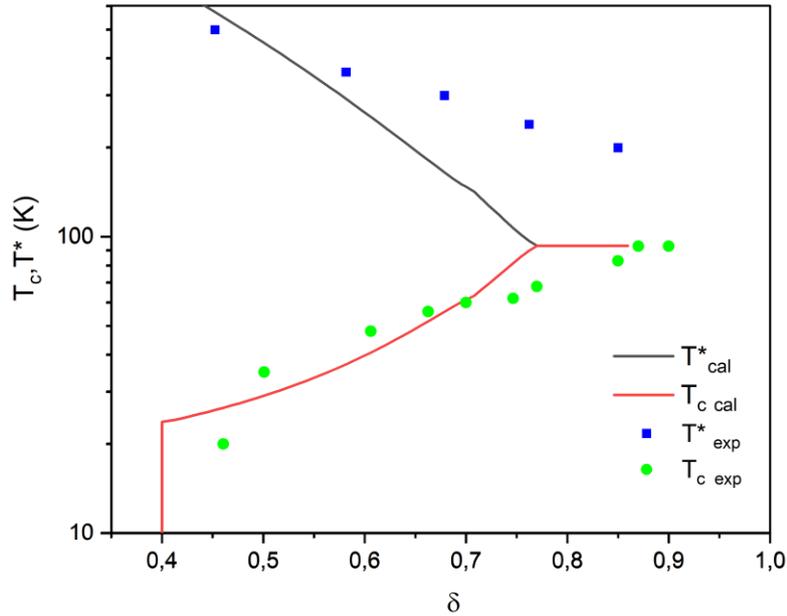

Fig. 7. Dependences of temperatures $T^*$ and $T_c$ in YBa$_2$Cu$_3$O$_{6+\delta}$ on doping. Solid curves are $T^*_{cal}(\delta)$ and $T_{c,cal}(\delta)$ calculated from (11), symbols are experimental values of $T^*_{exp}(\delta)$ and $T_{c,exp}(\delta)$ from [10] and [42], respectively. For $\delta<0.4$, $T^*_{cal}(\delta)=0$ because the size of the average cluster at these concentrations is $\bar{S}<2$, i.e. the formation of HL centers in these clusters is impossible.

It can be seen from the fig. 7 that the dependence $T_{c,cal}(\delta)$ satisfactorily describes the experimental dependence $T_{c,exp}(\delta)$, despite the use of the average cluster size for calculations. The sharp drop in $T_{c,exp}(\delta)$ to 0 at $\delta<0.4$ is explained by a decrease in the size of the average cluster at these concentrations to $\bar{S}$<2, which sharply reduces the concentration of HL centers and makes the formation of a Josephson-coupled superconducting cluster impossible. The smooth transition to the 93K phase at δ>0.77 (against the expected sharp percolation transition) can be explained by the presence of defects, a slight deviation of the composition from the stoichiometric, as well as partial filling of O5 sites in the plane of chains at δ→1.

At first sight, the calculated dependence $T^*_{cal}(\delta)$ significantly diverges from $T^*_{exp}(\delta)$, especially in the region δ ~1. However, these are exactly the differences that should be expected when the average cluster size $\bar{S}(\delta)$ is used to calculate $T^*_{cal}(\delta)$. Since small clusters that determine the temperature T* also exist at δ ~1 (when $\bar{S}(\delta)$→∞), the difference between $T^*_{cal}(\delta)$ and $T^*_{exp}(\delta)$ in the region of large δ is expected. On the other hand, at low concentrations, when the average cluster size $\bar{S}2(\delta)$ is small enough, $T^*_{cal}(\delta)$ is quite close to $T^*_{exp}(\delta)$.

## Other anomalies near $T_c$

As it follows from (10), as the temperature decreases from *T\** to *T*$_c$, the average occupation of HL centers decreases and, accordingly, the number of clusters being simultaneously in a superconducting state increases as well as their lifetime in this state. When approaching $T_c$, quite large clusters begin to switch to the superconducting state, for which the relative fluctuations in the population of HL centers are small, and their average lifetime in the superconducting state is relatively long. This makes it possible to observe above $T_c$ such phenomena as the anomalous Nernst effect [48] and giant diamagnetism [49].

Below Tc, a superconducting percolation cluster is formed from isolated clusters connected by Josephson bonds. Since isolated clusters can fluctuationally switch from the superconducting state to the normal state and back, these switchings will be accompanied by changes in the topology of the percolation cluster. As a result, the magnetic flux will be able to leave the sample, which will lead to the reversibility of the magnetization curves [50].

## Conclusion

In this work, we have shown that a number of unusual properties distinguishing underdoped cuprates from conventional superconductors appear to be of a common nature and result from the cluster structure of the underdoped phase and the specific mechanism of superconducting pairing in cuprates. This pairing mechanism is assumed to be mediated by the virtual scattering of electrons into unoccupied localized pair states formed on pairs of neighboring Cu cations under doping and is reminiscent of the Little-Ginsburg mechanism.

The cluster structure of the underdoped phase in combination with the special mechanism of superconductivity of cuprates leads to the fact that in a certain temperature range Tc<T<T* finite isolated clusters can exist in both superconducting and normal states, switching between these states randomly due to fluctuations of pair states occupation with electrons.

In this case, we can consider that below $T_c$ the cluster is almost always in a superconducting state, and above T* - in a normal state, and the interval $T_c$<T<T* can be identified with the region of existence of the so-called pseudogap phase. The temperatures $T_c$ and T* for $YBa_2Cu_3O_{6+\delta}$ were calculated depending on the doping level. The calculation results are in good agreement with experiment without any fitting parameters.

In the same temperature range, the contribution to optical conductivity from a sequence of randomly arising superfluid density pulses from each cluster can be represented as a random process with a correlation time $\tau_{corr}$~$10^{-15}$ sec, which corresponds to the effective spectrum width $\omega_{eff}$~1 eV and explains the seeming effect SW transfer from high-frequency region.

In the vicinity of $T_c$, fluctuation induced s-n switching of finite clusters are apparently the cause of another group of phenomena: the anomalous Nernst effect, giant diamagnetism (above $T_c$) and reversibility of magnetization curves (below $T_c$).

Thus, the approach under consideration allows us to explain many anomalous properties of underdoped cuprates. This fact can be considered as an indirect confirmation of the validity of the proposed model of HTSC cuprates.

**Authorship contribution statement**

**K. Mitsen**: Conceptualization, Model construction, Writing – original draft. **O. Ivanenko**: Development of this work, Supervision, Writing – review & editing.

**Declaration of competing interest**

The authors declare that they have no known competing financial interests or personal relationships that could have appeared to influence the work reported in this paper.

**Acknowledgments**